\input epsf
\documentstyle[preprint,eqsecnum,aps]{revtex}
\tightenlines

\def\INSERTFIG#1#2#3{
\multiply\baselineskip
by 3 \divide\baselineskip by 4
\vbox{\bigskip \vbox{\hfil\epsfbox{#1}\hfill}%
{
\noindent%
{FIG.\ #2. }{\sl #3 \medskip}}
}\multiply\baselineskip by 4 \divide\baselineskip by 3 \hfill}%

\begin{document}

\draft
\preprint{\vbox{
\hbox{UCSD/PTH 97--20}
}}

\title{Explicit Quark-hadron Duality in Heavy-light
Meson Weak Decays in the 't~Hooft Model}

\author{Benjam\'{\i}n Grinstein\footnote{bgrinstein@ucsd.edu} and
Richard F. Lebed\footnote{rlebed@ucsd.edu}}

\address{Department of Physics,
University of California at San Diego, La Jolla, CA 92093}
\date{August 18, 1997}

\maketitle
\begin{abstract}
	We compute the nonleptonic weak decay width of a heavy-light
meson in 1+1 spacetime dimensions with a large number of QCD colors
(the 't~Hooft model) as a function of the heavy quark mass.  In this
limit, QCD is exactly soluble, and decay modes are dominated by
two-particle final states.  We compare the results to the tree-level
partonic decay width of the heavy quark in order to test quark-hadron
duality in this universe.  We find that this duality is surprisingly
well satisfied in the heavy quark limit, in that the difference
between the sum of exclusive partial widths and the tree-level
partonic width approaches a constant as $M\to\infty$, and the
deviation is well-fit by a small $1/M$ correction. We comment on the
meaning of this conclusion and its implications for the use of
quark-hadron duality in hadronic physics.
\end{abstract}



\narrowtext

\section{Introduction}

	Quark-hadron duality, in its most general form, is the notion
that certain rates for processes involving hadrons can be computed
simply as the underlying partonic rates\cite{Bloom}.  Duality allows
us to compute many quantities which would otherwise be hopelessly
difficult.  One common application of duality is to the nonleptonic
weak decays of heavy hadrons.  The lore is that, for large enough
heavy quark mass, duality holds in the computation of the hadronic
width.

	Several discrepancies between theory and experiment that have
recently received attention rely on quark-hadron duality.  Among them
are the significant difference between lifetimes of beauty baryons and
mesons\cite{Ric}, the overestimates of the $B$-meson semileptonic
branching fraction\cite{bigi2} and the average number of charm quarks
per $B$ decay\cite{bagan}.  Because the limit of experimental
knowledge about nonleptonic $B$ decays is rapidly expanding, such
issues are of great topical interest.

	But when is duality valid?  In many cases duality follows from
the Operator Product Expansion (OPE).  This is the case, for example,
for the rate of $e^+e^-\rightarrow\hbox{hadrons}$ and for the
semileptonic decay rates of heavy hadrons.  However, duality is
applied in many other cases, such as in hadronic widths of heavy
hadrons, for which there is no OPE.

	Reference~\cite{bigi1} proposes an OPE-like expansion in
inverse powers of the heavy quark mass~$M$, which not only
incorporates quark-hadron duality as the lowest term in the expansion,
but also organizes the corrections by inverse powers of $M$.  A main
result of that work is the claim that corrections first appear at
order $1/M^2$.  The question above can be reformulated as, ``Is an
OPE-like expansion like that of Ref.~\cite{bigi1} valid?''

	To investigate the validity of duality it is convenient to
work with a soluble model of strong interactions formulated as a
full-fledged field theory, so that one may test duality both in cases
with and without an OPE.  The 't~Hooft model\cite{tH}, large-$N_c$ QCD
in $1+1$ dimensions, is a good laboratory for this purpose.  It
contains an infinite spectrum of mesons composed of confined quarks,
realizes asymptotic freedom trivially, and inherits all the
phenomenological consequences of large-$N_c$ QCD\cite{tHN} common to
our universe, such as dominance of scattering amplitudes with the
minimum number of meson states, OZI suppression, the absence of
exotics, and others.  For processes with an OPE, duality in the 't
Hooft model has been checked explicitly\cite{CCG,Ein}.  However,
little is known about duality for non-OPE processes.  The reason is
that, in precisely those cases for which an OPE is lacking, there is
no simple analytical method of verifying duality, and one must resort
to arguments based on numerical solutions.

	In this paper we compute the hadronic weak decay width
$\Gamma(M)$ of heavy ``$B$'' mesons in the 't~Hooft model as a
function of the heavy ``$b$'' quark mass~$M$; the meaning of ``heavy''
is made precise in Sec.~\ref{review}.  We compare this to the partonic
(perturbative) decay width of the heavy quark, $\Gamma_{\rm part}(M)$,
which we compute analytically.  For large $M$ we find that both
$\Gamma_{\rm part}(M)$ and $\Gamma(M)$ are essentially linear in
$M$. The difference between the two appears to be asymptotically
constant and small, indicating a small $1/M$ correction to the naive
duality limit. As $M$ increases, new hadronic decay channels become
accessible, and at each of these thresholds there is a singular peak
in $\Gamma(M)$.  Averaging $\Gamma(M)$ over a region in $M$ that
includes many resonances removes these peaks but does not change the
leading dependence on $M$.  {\it Our conclusion is that duality holds
to leading order in $M$, but unlike the OPE-like expansion of
Ref.~\cite{bigi1}, appears to have $1/M$ corrections.}

	In Ref.~\cite{roma} it is argued that there is strong experimental
evidence for the failure of duality.  What is meant there is that the
pattern of corrections in powers of $1/M$ of Ref.~\cite{bigi1} is not
supported by experiment.  This agrees with our result, which indicates
a violation to duality at first order in the $1/M$ expansion of
Ref.~\cite{bigi1}.

	While we cannot carry over our quantitative results to the
physical world of non-planar QCD in $3+1$ dimensions, we believe that
there is nothing intrinsic to 1+1 dimensions that would make duality
work differently than in 3+1.  The operator analysis that leads to the
$1/M$ expansion proceeds in 1+1 much as in 3+1.
	
	The paper is organized as follows.  In Sec.~\ref{review}, we
briefly review the 't~Hooft model and a standard method for its
numerical solution.  Section~\ref{1+1} compares features of 1+1
dimensions, such as the nature of phase space and spin, to those in
3+1 dimensions.  In Sec.~\ref{inc}, we present the algebraic results
of the inclusive parton-level calculation of widths.  In
Sec.~\ref{exc}, we present the results of the exclusive calculation in
the 't~Hooft model.  Section~\ref{res} gives our numerical results and
a discussion of their implications, and Sec.~\ref{conc} concludes.

	There are other nonperturbative questions of phenomenological
interest in the area of hadronic $B$ weak decays for which there is an
established lore.  One can test any of these hypotheses in the
't~Hooft model.  Of particular interest is the notion that
contributions to decay amplitudes from different underlying quark
diagram topologies contribute with distinct weights, which can very
much suppress the amplitude from a given topology.  For example, the
``annihilation'' diagrams, in which the valence quark-antiquark pair
annihilate through a weak current, are supposedly suppressed relative
the ``spectator'' diagrams by a factor of $f_B/M_B$, where $f_B$ is
the $B$ decay constant and $M_B$ its mass.  We will address this
question in a separate publication\cite{GLII}.

\section{The 't~Hooft Model} \label{review}

	The success of 't~Hooft's method of solving a strongly-coupled
theory rests on two assumptions that considerably simplify the
problem.  First, one works in the limit of large $N_c$, in which it is
readily seen\cite{tHN} that diagrams including either internal
fermion-antifermion loops or the crossing of gluon lines at points
other than their vertices are suppressed by combinatorial powers of
$N_c$ compared to those that do not.  These simple topological
consequences of the theory lead directly to the predictive power of
large $N_c$.  Second, in 1+1 dimensions one may use the gauge freedom
of QCD to choose a linear gauge in which some chosen component of the
gluon field vanishes, and only the sole orthogonal component survives.
Then, since the gluon self-coupling term in the field strength appears
as a commutator of field components, this term vanishes in the gauge
we have selected.  Consequently, gluon self-coupling vanishes in this
gauge, and so in combination with large $N_c$, gluon lines are not
permitted to cross each other, even at vertices.  Moreover, ghosts are
absent in linear gauges.  It follows that the only diagrams that must
be summed are ``rainbow'' diagrams for the quark mass and wave
function renormalization, and ``ladder'' diagrams for quark-antiquark
interactions\cite{tH}.

	In 1+1 dimensions confinement is realized trivially, since the
lowest-order inter-quark potential obtained by taking the Fourier
transform of the gluon propagator (which gives rise to the $1/r$
Coulomb interaction in four dimensions) grows linearly with the
inter-quark separation.  Although lowest-order color confinement is an
automatic consequence in two dimensions, it is a highly nontrivial
fact that the phenomenon persists in the all-orders Green function
solutions of the 't~Hooft model.

	To be specific, the Lagrangian of QCD, as in four dimensions,
is
\begin{equation}
{\cal L} = -\frac 1 4 {\rm Tr} \, F_{\mu \nu} F^{\mu \nu} + \sum_a
\bar \psi_a \left( \gamma^\mu (i \partial_\mu - g A_\mu) -m_a \right)
\psi_a ,
\end{equation}
where $A_\mu$ is the $SU(N_c)$ gauge field with field strength $F_{\mu
\nu}$ defined in the usual way, and $\psi_a$ is a Dirac fermion of
bare mass $m_a$ and flavor $a$.  The bare coupling $g$ not only has
dimensions of mass in two dimensions, but scales as $1/\sqrt{N_c}$ in
the large-$N_c$ limit.

	The renormalization of the fermion propagator is exceptionally
simple.  The only modification is a shift of the bare fermion mass by
\begin{equation} \label{mren}
m_a^2 \to m_{a,R}^2 \equiv m_a^2 - g^2 N_c / 2 \pi .
\end{equation}
Consequently, it makes good sense to describe masses in units of $g
\sqrt{N_c / 2 \pi}$, which is finite in the $N_c \to \infty$ limit.
The dividing line of $m_a^2 = 1$ ($m_{a,R}^2 = 0$) in these units acts
as a boundary between heavy and light quarks, as is numerically
verified in Refs.~\cite{JM,GM1,GM2}; for example, in \cite{GM1} it was
seen that the meson decay constant approaches the standard asymptotic
behavior $f_B \propto 1/\sqrt{M}$ for $M \geq 5$ or so.  It follows
that $g\sqrt{N_c / 2 \pi}$ in 1+1 assumes a role analogous to
$\Lambda_{\rm QCD}$ in 3+1.

	Quantization of the theory is most convenient in axial
light-cone gauge ($A_- = 0$), where light cone coordinates are defined
by
\begin{equation}
x^{\pm} \equiv x_{\mp} \equiv \frac{(x^0 \pm x^1)}{\sqrt{2}} ,
\end{equation}
and analogously for other vectors.  The chief advantage of this choice
is that only one component of the Dirac algebra ($\gamma_-$) survives,
thus effectively eliminating the need to perform Dirac traces.

	Upon solving for the Green function of a fermion-antifermion
pair with bare masses $M$ and $m$ in this model, one obtains the
bound-state eigenvalue equation
\begin{equation} \label{tHe}
\mu_n^2 \phi_n^{M\overline{m}} (x) = \left( \frac{M_R^2}{x} +
\frac{m_R^2}{1-x} \right) \phi_n^{M\overline{m}}(x) - \int^1_0 dy \,
\phi_n^{M\overline{m}} (y) \, \Pr \frac{1}{(y-x)^2},
\end{equation}
which is known as the 't~Hooft equation\cite{tH}.  Here the $n$th
eigenstate $\phi_n^{M\overline{m}}$ is the meson wave function, the
$n$th eigenvalue $\mu_n^2$ is its squared mass, and $x$ is the
fraction of the meson momentum's minus component (which acts, in the
light-cone quantization, as a canonical spatial momentum component)
carried by quark $M$.  We will always label the ground state (the
lowest mass meson) by $n=0$.  The principal value prescription serves
to regulate the integrand singularity, which originates in the
infrared divergence of the gluon propagator.  This equation has a
discrete spectrum of eigenvalues that increase approximately linearly
for large $n$, and the wave functions vanish at the boundaries $x=0$
and 1, with the asymptotic behavior $\phi_n^{M\overline{m}} (x) \to
x^{\beta_M}$ as $x
\to 0$, where
\begin{equation} \label{asymp}
M^2_R + \pi \beta_M \cot \pi \beta_M = 0 ,
\end{equation}
and similarly as $x \to 1$, exchanging $m$ for $M$.  $\beta_M$ is a
monotonic function of $M^2$ (or $M_R^2$), increasing from zero to one
as $M^2 = 0 \to \infty$.

	Also useful in this description is the full meson-quark vertex
$\Phi_n^{M\overline{m}}$, which is given by
\begin{equation} \label{bphi}
\Phi_n^{M\overline{m}} (z) = \int_0^1 dy \, \phi_n^{M\overline{m}} (y)
\, \Pr \frac{1}{(y-z)^2} ,
\end{equation}
for all complex values of $z$.  Indeed, except for $z \in$ [0,1], the
principal value prescription is unnecessary.

	The decay constant $f_n$ for meson $n$ may be computed in this
framework.  It is given by
\begin{equation} \label{dec}
f_n = c_n / \sqrt{\pi} ,
\end{equation}
where
\begin{equation} \label{cn}
c_n \equiv \int_0^1 dx \, \phi_n(x) .
\end{equation}
Strictly speaking, the r.h.s.\ of (\ref{dec}) is also multiplied by a
factor $\sqrt{N_c}$, but we may absorb this factor into the
normalization of other factors by which it is multiplied in the full
amplitudes; what is important is that the final physical amplitude has
the correct $N_c$ dependence at leading order.  As each new quantity
is calculated in this paper, we will point out the leading dependence
on $N_c$, but as a rule we suppress the explicit factors for ease of
notation.  The 't~Hooft eigenfunctions $\phi_n$, for example, are
$O(N_c^0)$ solutions of Eq.~(\ref{tHe}), and so $c_n$ is also
$O(N_c^0)$.  On the other hand, light meson decay constants have the
well-known behavior $f_n \propto \sqrt{N_c} \,$, and the full result
Eq.~(\ref{dec}), including the $\sqrt{N_c} \, $, may be verified by
direct calculation.

	The 't~Hooft model wave functions $\phi_n$ and $\Phi_n$ are
calculated by means of a standard numerical method called the Multhopp
technique\cite{JM,Mul}, in which the integral equation is converted to
an equivalent infinite-dimensional eigenvalue system, which in turn
may be truncated after a desired number of modes to give approximate
wave function solutions.  Since the relevant formul{\ae} for
unequal-mass mesons do not appear elsewhere, we present a summary in
Appendix A.

	Whereas the eigenfunctions $\phi_n^{M\overline{m}}$ describe
the complete set of homogeneous solutions for the two-point Green
function, the solution for $1 \to 2$ meson decays (the leading decay
channels in large $N_c$) requires also three-point Green functions.
Remarkably, the requisite expressions may be written entirely in terms
of triple overlap integrals of the functions $\phi$ and
$\Phi$\cite{Ein,GM2}, without bare quark model contact-type
interactions.  In physical terms, this means that the three vertices
of the diagram for the three-point Green function are resonance
dominated, without contact contributions.  Nevertheless, for the
diagrams computed, it will prove to be computationally convenient to
describe part of the full amplitude in terms of these contact terms.
We exhibit these explicit expressions in Sec.~\ref{exc}, but for the
moment it is only important to note that such expressions indeed
exist.

\section{Peculiarities of 1+1 Dimensions} \label{1+1}

	Despite one's hopes that exact calculations in the 't~Hooft
model may lend insight into real 3+1 strong interaction physics, we
emphasize that the two-dimensional universe possesses some unique
properties that must be remembered when comparing to the universe of
four dimensions.  Therefore, the 't~Hooft model may in no way be
construed as any sort of limiting case of real QCD, and any direct
comparisons are necessarily qualitative.  In other words, we espouse
the opinion that only certain conclusions based upon our numerical
studies of 't~Hooft model solutions, not the numerical results
themselves, possess any validity in 3+1 dimensions.

	The most obvious signal that 1+1 and 3+1 physics are vastly
different is that the former does not possess the quantity of angular
momentum, except in the residual form of parity\footnote{Also, spinors
retain the property of chirality, since a $\gamma_5$ matrix still
exists in 1+1, signaling two inequivalent representations of the
Lorentz group.}.  This is clear since finite rotations do not exist
when there is only one spatial direction, and only the improper
``rotation'' taking $x^1 \to -x^1$, namely parity, remains.  It
follows that 't~Hooft model eigenstates $\phi_n(x)$ do not possess
spin, but only intrinsic parity $(-1)^{n+1}$\cite{CCG}.  All of the
interesting phenomenology provided by approximate spin symmetries in
our world ({\it i.e.}, the smallness of hyperfine splittings,
relations between different helicity amplitudes, {\it etc.}) are
therefore meaningless in two dimensions.

	The lack of transverse directions has important consequences
for couplings in 1+1 dimensions.  As mentioned in the previous
section, gauge couplings have dimensions of mass, and so such theories
are super-renormalizable.  Moreover, ``vector'' gauge bosons exist in
1+1 only through their longitudinal modes.  There are also different
Lorentz invariants in 1+1, since the Levi-Civita tensor
$\epsilon^{\mu\nu}$ has only two indices.  The effects of these
constraints are implicit in all the results to follow.

	The amount of Lorentz-invariant phase space is of course
expected to vary between different spacetime dimensions $D$, since the
measure of the phase space integrals is the $D$-dimensional volume
element.  However, the difference between 1+1 and 3+1 is particularly
dramatic.  To be specific, in $D$ spacetime dimensions, the
differential width for a $1 \to 2$ decay in terms of the solid angle
of either final-state particle is given by
\begin{equation} \label{two}
d\Gamma = \frac{|{\bf p}|^{D-3}}{(2\pi)^{D-2} \, 8M^2} |{\cal
M}|^2 d\Omega,
\end{equation}
where $|{\bf p}|$ is the spatial momentum of either final-state
particle in the rest frame of the initial particle of mass $M$, and
${\cal M}$ is the invariant amplitude of the process.  Note
particularly the behavior of the phase space factor $|{\bf p}|^{D-3}$
as the $|{\bf p}| = 0$ threshold is approached: For $D=4$, the
differential width vanishes with the decreasing amount of phase space
available, but for $D=2$, the differential width actually becomes
singular (barring an accidental zero in the amplitude ${\cal M}$).  It
follows that two-particle decay modes near threshold are enhanced in
1+1, in stark contrast to 3+1.

\section{The Partonic Calculation} \label{inc}

	Because of the small number of integrations necessary to
compute phase space in 1+1 dimensions, it is possible to perform the
partonic integrals analytically in all cases of interest to us.  In
the $1 \to 3$ parton decay, one starts with $3 \times 2 = 6$
final-state momentum components, of which 2 are fixed by
energy-momentum conservation and 3 are fixed by the on-shell
conditions of the final-state partons; this leaves only one nontrivial
integration, which can be done explicitly.\footnote{Strictly speaking,
there is also a degree of freedom from the ``solid angle'' of one of
the final-state particles.  However, in 1+1, this is a discrete degree
of freedom.  Integration of a differential width over this quantity
gives an additional factor of $1+1=2$ for Lorentz scalars and
$1+(-1)=0$ for pseudoscalars.}

	Here we consider the case of an initial quark of mass $M$
decaying into three distinguishable equal-mass quarks of mass $m \leq
M/3$.  The final nontrivial integral involves a small number of square
root factors arising from the on-shell mass-energy constraints, since
both energies and momenta appear in both the phase space and the
invariant amplitude expressed in a given frame.  Such expressions in
our case integrate to the standard three kinds of complete elliptic
integrals of Legendre, usually denoted by $K$, $E$, and $\Pi$.  We
begin by presenting the functional form for the phase space with
constant invariant amplitude:
\begin{equation}
\Phi_3 (M;m,m,m) = \frac{1}{4\pi^3 M^2} (1 + \epsilon)^{-1/2} \left( 1
- \epsilon/3 \right)^{-3/2} K (u) ,
\end{equation}
where
\begin{equation} \label{eps}
\epsilon \equiv \frac{3m}{M} \in [0,1] ,
\end{equation}
and
\begin{equation} \label{u}
u \equiv \sqrt{\frac{(1-\epsilon)(1+\epsilon/3)^3}
{(1+\epsilon)(1-\epsilon/3)^3}} .
\end{equation}
Note that, unlike the two-body phase space given by Eq.~(\ref{two}),
this expression does not diverge for finite $m$.  However, it does
possess a singularity as $m \to 0$ ($\epsilon \to 0$),
since then
\begin{equation}
\Phi_3 = \frac{3}{8\pi^3 M^2} \ln \left( \frac{M}{m} \right) \left[ 1
+ O \left( \frac{m^2}{M^2} \right) \right] .
\end{equation}
The opposite limiting case $\epsilon \to 1$, in which the three
partons are produced at rest, is equally peculiar:
\begin{equation}
\Phi_3 = \frac{3 \sqrt{3}}{32 \pi^2 M^2} \left[ 1 + O(1-\epsilon)
\right] ,
\end{equation}
which means that phase space does not vanish in this limit.

	We now present the expressions for the inclusive partonic
decay width.  For definiteness, we attempt to describe the couplings
in terms as similar to Standard Model (SM) notation as possible.  Our
labeling of partons is exhibited in Fig.~1.  The decay of the heavy
quark 1 to the lighter quark 3 is assumed to couple to a vector-like
weak current with vertex factor $(-ig_2/\sqrt{2}) V_{31}
\gamma^\mu \left( c_V - c_A \gamma_5 \right)$, carried by a gauge
boson ``$W$'' of mass $M_W$; in the SM, $c_V = c_A = 1/2$.  The
coupling at the other end of the weak current, creating quark 5 and
antiquark 4, is assumed to be the same except for the ``CKM element''
$V^*_{45}$.  $G_F$ is defined, as in the SM, by $\sqrt{2} g_2^2 /
8M_W^2$; note that $G_F$ is dimensionless in 1+1.  Finally, the
abbreviations $\epsilon$ and $u$ are carried over from
Eqs.~(\ref{eps}) and (\ref{u}).

\INSERTFIG{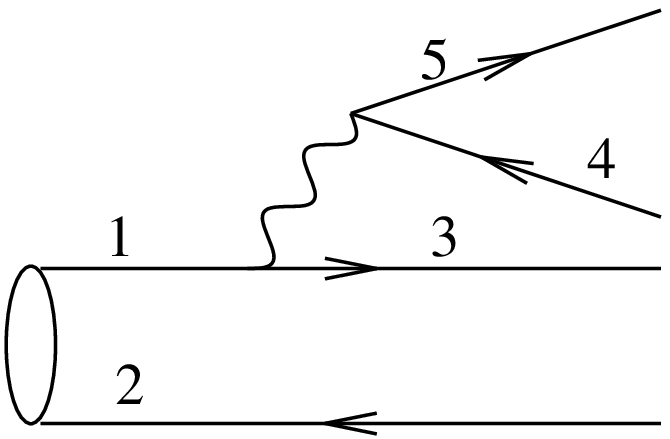}{1}{Parton decay diagram for the inclusive decay.
Of interest are the parton labels, as used in the text.}

	The weak decay amplitude and width in this case are effects of
orders $\sqrt{N_c}$ and $N_c$, respectively.  This counting may be
established in the parton diagram by observing that the pair ($5\bar
4$) in Fig.~1 can occur with each of the $N_c$ colors, but ``sewing
up'' the partons into color-singlet mesons ($5\bar 4$), ($1\bar 2$),
and ($3\bar 2$) costs a factor of $1/\sqrt{N_c}$ each.  Finally, each
color may occur in the loop created by 1, 3, and 2, for one more
factor of $N_c$.  It follows that the weak decay width\footnote{The
difference from the strong width, which is $O(1/N_c)$, is that the
$q\bar q W$ vertices are unsuppressed in large $N_c$, while the $q\bar
q$-gluon vertex is $O(1/\sqrt{N_c}$).}  calculated from the parton
diagram is $O(N_c)$.  In Sec.~\ref{exc} we show that the hadronic
calculation of the width also produces a leading factor of $N_c$.

	The width is presented in two special cases.  In the first, we
take $M_W \gg M$, the usual four-fermion coupling assumption.  This
corresponds to using only the $g_{\mu\nu}$ term in the numerator of
the $W$ propagator\footnote{We use unitary gauge in order to avoid the
necessity of including additional charged Goldstone fields.}, $-i
\left( g_{\mu\nu} - q_\mu q_\nu /M_W^2 \right) /(q^2+i\varepsilon)$.
We then find
\begin{eqnarray} \label{partw1}
\Gamma & = & \frac{4G_F^2 M}{\pi} |V_{31}V_{45}^*|^2 (c_V^2 - c_A^2)^2
\left( 1 - \epsilon/3 \right)^{3/2} ( 1 + \epsilon)^{1/2} \nonumber \\
& & \times \left[ E(u) - 16 \left( \epsilon / 3 \right)^3 \left( 1 -
\epsilon/3 \right)^{-3} ( 1 + \epsilon)^{-1} K(u) \right] .
\end{eqnarray}
The limiting cases of this expression are given by
\begin{equation} \label{lim1}
\Gamma \to \frac{4G_F^2 M}{\pi} |V_{31}V_{45}^*|^2 (c_V^2 - c_A^2)^2
\left[ 1 - \frac{\epsilon^2}{3} + O(\epsilon^3 \ln \epsilon) \right] ,
\end{equation}
as $\epsilon \to 0$, and
\begin{equation}
\Gamma \to \frac{16G_F^2 M}{3\sqrt{3}} |V_{31}V_{45}^*|^2
(c_V^2 - c_A^2)^2 (1-\epsilon) \left[ 1 + \frac{3}{4} (1-\epsilon) +
O((1-\epsilon)^2) \right],
\end{equation}
as $\epsilon \to 1$.
We see that the width is finite as $\epsilon \to 0$ and vanishes as
$\epsilon \to 1$.  The former limit shows that the width for the
partonic decay of the heavy quark is given approximately by $M$ times
a constant, dimensionless coefficient.

	It is easy to understand the prefactor $(c_V^2 - c_A^2)^2$,
which means that the width in the $M_W \gg M$ limit vanishes for $V
\pm A$ currents.  The decay vertex in this limit is of the form
$g^{\mu\nu}J_\mu j_\nu$ for some quark currents $J$ and $j$.  Note
that the only non-vanishing components of the metric are $g^{+-}$ and
$g^{-+}$, so the vertex involves only $J_-j_+$ and $J_+j_-$.  Now, $V
\pm A$ currents correspond to the quarks being all right(left)-handed.
The currents $J_{\pm}$ and $j_{\pm}$ in this chiral basis are just
bilinears of $\gamma_{\pm}$, since $\gamma_\mu \gamma_5 \psi_{R,L} =
\pm \gamma_\mu \psi_{R,L}$.  However, in $1+1$ dimensions $\gamma_-
\psi_L=\gamma_+ \psi_R=0$, and so all currents of one chirality
vanish, leading to the vanishing of the decay vertex.

	If we impose $c_V^2 = c_A^2 = 1/4$ from the beginning of the
calculation, then we find that only pieces obtained from contraction
with the $q_\mu q_\nu$ terms in the $W$ propagator survive, giving
rise to the width
\begin{eqnarray} \label{partw2}
\Gamma & = & \frac{G_F^2 M^5}{\pi M_W^4} |V_{31} V^*_{45}|^2
\left( \epsilon / 3 \right)^2 \left( 1 - \epsilon /3 \right)^{-3/2}
(1 + \epsilon)^{-1/2} \nonumber \\
& & \times \Biggl\{ -8 \left( \epsilon / 3 \right)^2 \left[ \left( 1+
\epsilon /3 \right)^3 + (\epsilon/3) \left( 1 - \epsilon /3 \right)^2
\right] K (u) \Biggr. \nonumber \\
& & + \left( 1 + \epsilon^2 / 9 \right)
\left( 1 - \epsilon / 3
\right)^3 ( 1 + \epsilon ) E(u) \nonumber \\ & & \Biggl. + 48 (
\epsilon / 3)^3 \left( 1 + \epsilon^2 / 27 \right) \Pi (v,u)
\Biggr\} ,
\end{eqnarray}
where
\begin{equation}
v \equiv
\frac{(1+\epsilon/3)(1-\epsilon)}{(1-\epsilon/3)(1+\epsilon)}.
\end{equation}
The asymptotic expansions in this case are
\begin{equation}
\Gamma \to \frac{G_F^2 M^5}{\pi M_W^4} \left(
\frac{\epsilon}{3} \right)^2 \left[ 1 - \frac{2\epsilon^2}{9} +
O(\epsilon^3) \right] ,
\end{equation}
for $\epsilon \to 0$, and
\begin{equation}
\Gamma \to \frac{8 G_F^2 M^5}{243 \sqrt{3} M_W^4} \left[ 1 +
\frac{1}{2} (1-\epsilon) + O((1-\epsilon)^2) \right],
\end{equation}
for $\epsilon \to 1$.  Here we see that the width vanishes as
$\epsilon \to 0$ but is finite as $\epsilon \to 1$.  Specifically, in
the former limit the width is approximately a dimensionless constant
times $M^3 m^2/M_W^4$.

\section{The Hadronic Calculation} \label{exc}

	Matrix elements for exclusive $1 \to 2$ meson decays are most
conveniently written in terms of transition form factors.  We identify
the $\bar B$ meson in 1+1 as the ground state of the $M \overline{m}$
tower of resonances to which it belongs, and subsequently label it by
{\bf 0}.  Consider the ``tree'' (T) diagram of
Fig.~2, for which $\bar B = (1 \bar 2)$, where 1 is the heavy ``$b$''
quark; the matrix element is parameterized by
\begin{equation} \label{mat}
\langle {\bf m} (p') \left| \bar q \gamma^\mu Q \right| {\bf 0} (p)
\rangle = \left\{ \begin{array}{ll}
(p+p')^\mu f_+ (q^2) +(p-p')^\mu f_- (q^2) & \hbox{for $m$ even,} \\
\epsilon_{\mu\nu} (p+p')^\nu f_+ (q^2) +
\epsilon_{\mu\nu} (p-p')^\nu f_- (q^2) & \hbox{for $m$ odd,} 
\end{array} \right.
\end{equation}
where $q^2 \equiv (p-p')^2$, and $Q$ and $q$ indicate the fields of
quarks with masses $M$ and $m$.  The light quark field $q$ here refers
to the daughter of the heavy quark (3 in Fig.~2), not the spectator
quark (2 in Fig.~2), although both are taken to have mass $m$.  The
label {\bf m} indicates the eigenvalue index of the final-state decay
product meson ($2\bar 3$) not coupled to the flavor-changing current.
In the remainder of this section, $m$ exclusively means this value and
not the value of the light quark mass.

\INSERTFIG{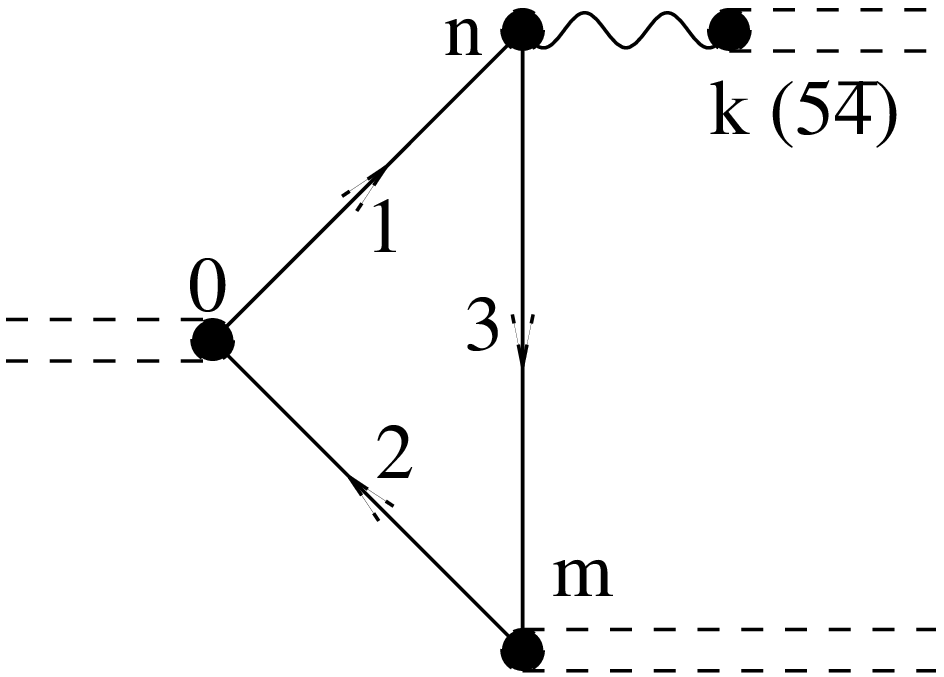}{2}{Diagram for ``tree'' (T) meson exclusive
decay.  Numbers indicate quark labels used in the text (except {\bf
0}, which refers to the ground-state ``$\bar B$'' meson), while
letters indicate the eigenvalue index of meson resonances.  One can
also consider contact-type diagrams, in which the point labeled by\/
{\bf n} is not coupled to a resonance.}

	Next, we reserve the label {\bf n} in the T diagram for the
meson resonances or contact terms ($1 \bar 3$) coupled to the
flavor-changing current.  {\bf n} carries the momentum transfer $q^2$,
which is the kinematic variable of interest in this system; however,
it proves more convenient to use the equivalent Lorentz-invariant
quantity $\omega \equiv p_-/q_-$, which indicates the fraction of
light-cone coordinate ``spatial'' component of the current $q_-$
carried by meson {\bf 0}.  In the method of calculating the matrix
element (\ref{mat}) developed in \cite{JM}, one considers not ${\bf 0}
\to {\bf mn}$ directly, but rather the crossed process ${\bf n} \to
{\bf 0} \overline{\bf m}$ above its threshold ($q^2 \geq (\mu_0 +
\mu_m)^2$); in that case, one finds $\omega \in [0,1]$:
\begin{equation} \label{om1}
\omega (q^2) = \frac{1}{2} \left[ 1 + \left( \frac{\mu_0^2 -
\mu_m^2}{q^2} \right) - \sqrt{1- 2
\left( \frac{\mu_0^2 + \mu_m^2}{q^2} \right) + \left( \frac{\mu_0^2 
- \mu_m^2}{q^2} \right)^2} \, \right].
\end{equation}
Here and below we use the same symbol $\mu$ for the masses of
heavy-light and light-light mesons, since from the index one can
immediately tell which one is appropriate ({\it e.g.}, $\mu_0$ is
heavy-light).  Since $\omega$ is obtained by solving the quadratic
equation $q^2 \omega^2 + (\mu_m^2 - \mu_0^2 -q^2) \omega + \mu_0^2 =
0$, it should be pointed out that the branch choice used for $\omega$
does not affect the final numerical results for form factors or
amplitudes; the two branches simply correspond to the two possible
directions of the mesons {\bf 0} and $\overline {\bf m}$ in the rest
frame of {\bf n}.  However, the branch chosen above turns out to
greatly facilitate the numerical computations.  For some values of
$q^2$ below this crossed-process threshold, $\omega$ is complex, and
the following expressions for the form factors must be computed in a
different way, as discussed below.

	With the aforementioned identifications, we may express the
form factors entirely in terms of resonance quantities, as promised in
Sec.~\ref{review}.  The notation and form factor expressions we
present here appear in Ref.~\cite{GM2}, while the characteristic
integral expression contained within was first obtained in~\cite{Ein}.
The form factors are explicitly
\begin{equation} \label{fp1}
f_+ (q^2) = \sum_n \frac{{\cal A}_n (q^2)}{1-q^2/\mu_n^2},
\end{equation}
and
\begin{equation} \label{fm1}
f_- (q^2) = \frac{1}{q^2} \left[ \sum_n \frac{{\cal
B}_n (q^2)}{1-q^2/\mu_n^2} - \sum_n \frac{{\cal A}_n
(q^2)}{1-q^2/\mu_n^2} \left( \mu_0^2 - \mu_m^2 \right) \right] ,
\end{equation}
where the pole residue functions ${\cal A}_n$ and ${\cal B}_n$ are
given by
\begin{equation} \label{an1}
{\cal A}_n (q^2) =  \frac{c_n \left[ 1 + (-1)^{n+m} \right]}
{(q^2 \omega - \mu_0^2/\omega)} F_{n0m} (\omega) ,
\end{equation}
and
\begin{equation} \label{bn1}
{\cal B}_n (q^2) =  c_n \left[ 1 + (-1)^{n+m+1} \right] F_{n0m}
(\omega) ,
\end{equation}
and the triple overlap integral $F_{n0m}$ is defined by
\begin{eqnarray} \label{fn0m}
\lefteqn{F_{n0m} (\omega) \equiv} & & \nonumber \\ & &
\left[ \frac{1}{1-\omega} \int_0^\omega dv \,
\phi_n^{1\bar 3} (v) \phi_0^{1\bar 2} \left(\frac v \omega \right)
\Phi_m^{2 \bar 3} \left( \frac{v-\omega}{1-\omega} \right) - \frac 1
\omega \int^1_\omega dv \, \phi_n^{1\bar 3} (v) \Phi_0^{1 \bar 2}
\left( \frac v \omega \right) \phi_m^{2\bar 3} \left(
\frac{v-\omega}{1-\omega} \right) \right] .
\end{eqnarray}

	We now compute the invariant matrix element for a current
coupling of the form $(-ig_2/\sqrt{2}) V_{31} \gamma^\mu \left( c_V -
c_A \gamma_5 \right)$, exactly as for the inclusive decay.  Such a
calculation is possible for an arbitrary combination of $V$ and $A$
currents, even though we presented the matrix element only for a
current of the bilinear $V^\mu \equiv \bar q \gamma^\mu Q$, because
the two currents are related by $\gamma_\mu \gamma_5 =
\epsilon_{\mu\nu} \gamma^\nu$.  The invariant matrix element is simply
the product of a linear combination of the form factors determined by
the current coupling, multiplied by the propagator of the
flavor-changing current (the ``$W$'') and finally by a factor
representing the meson formed from the flavor-changing current.  The
last step amounts, via LSZ reduction of the two-point Green function,
to the insertion of a factor of the meson decay constant.  In the T
diagram, we assign this meson the label {\bf k} and quark structure
($5 \bar 4$).  Using Eq.~(\ref{dec}) to write the decay constant
$f_{k}$ in terms of $c_{k}$, we have at last
\begin{eqnarray} \label{mt1}
{\cal M}_T & = & c_{k} \sqrt{\frac{2}{\pi}} \frac{G_F M_W^2}
{(M_W^2-q^2)} V_{31} V^*_{45} \nonumber \\
& & \cdot \sum_n \Biggl\{ 2\left[ (c_V^2-c_A^2) \left( (-1)^{k} +
(-1)^n \right) \right] \Biggr. \nonumber \\
& & \Biggl. \hspace{2em} + \frac{q^2}{M_W^2} \left[ (c_V + c_A)
(-1)^{k} - (c_V - c_A) \right] \left[ (c_V + c_A) (-1)^n - (c_V -
c_A) \right] \Biggr\} \nonumber \\
& & \cdot \frac{c_n \mu_n^2}{q^2 - \mu_n^2} F_{n0m} (\omega) ,
\end{eqnarray}
where the on-shell process has $q^2 = \mu_{k}^2$.  The pseudoscalar
parity of the ground state $\bar B$ has been taken into account in
this expression.  We remind the reader that in this expression $
F_{n0m}$ is given by (\ref{fn0m}) only for $n$ such that $\mu_n^2 >
(\mu_0 + \mu_m)^2$; other methods must be employed for smaller
$\mu_n^2$, as described below.

	The conversion of the decay constant $f_{k}$ to $c_{k}$ in
fact gives the only surviving factor of $\sqrt{N_c}$ in the amplitude,
which means that the weak decay width is proportional to $N_c \,$, in
agreement with the partonic result of Sec.~\ref{inc}.  This may be
seen with reference to Fig.~2 by the usual large $N_c$ counting
arguments: The coupling of three mesons ({\bf 0}, {\bf m}, and {\bf n}
in this case) appears with the factor $1/\sqrt{N_c}$, while meson {\bf
n} is destroyed by the $W$ current, thus providing a decay constant at
$O(\sqrt{N_c})$.  This part of the diagram alone, which is none other
than the form factors $f_{\pm} (q^2)$, is thus $O(N_c^0)$.  The
creation of the meson {\bf k} from the weak current gives the
remaining factor of $\sqrt{N_c}$.  Finally, the width is given by
Eq.~(\ref{two}).

	The question remains what to do with the contributions from
current-coupled resonances below the thresholds for the T diagram.
The expressions listed above are inadequate because they assume the
reality of $\omega$ in the computation of contour integrals with
denominators of the form ($\omega + c \pm i\varepsilon$), where the
$\varepsilon$ arises from the Feynman prescription in the fermion
propagators, and $c$ represents other purely real numbers arising from
the loop calculation.  Such integrals are naturally trivial when
$\omega$ is real and simply lead to step functions.  However, when
$\omega$ is complex the results are rather more cumbersome (although
still tractable in principle).  One may resort instead to other
methods\cite{JM} in order to obtain amplitudes from the
below-threshold resonances.  The approach relies on sum
rules\cite{Ein} that are satisfied by the amplitudes, and are
described in Appendix B\@.  The upshot is that the sum rules may be
used to describe the below-threshold amplitudes in terms of
combinations of values from the above-threshold amplitudes, a process
to which we refer as ``backsolving''.

	However, backsolving has drawbacks from the practical point of
view.  As the number of resonances below threshold increases, the
number of the above-threshold pole residues and corresponding
accuracies with which these are computed must increase dramatically to
maintain the accuracy of the below-threshold residues thus calculated.
For transitions of the $\bar B$ meson to light $\pi$'s, it is
known\cite{JM,GM2} that the below-threshold pole residues are very
large compared to their above-threshold fellows, and tend to alternate
in sign.  Clearly, a small uncertainty in the above-threshold
calculation magnifies to a large uncertainty in the below-threshold
residues, and the alternating sign suggests delicate cancellations
among the computed residues, which makes the situation even worse.

	There is a much more efficient method of calculation if we are
willing to abandon the requirement that all vertices in the
calculation of Fig.~2 are resonant couplings, and allow for quark
model-type contact terms.  The calculation is based upon the
observation that $\omega$ as defined in Eq.~(\ref{om1}) is real not
only for decays in the crossed kinematic region $q^2 \geq (\mu_0 +
\mu_m)^2$, but also decays in the direct decay kinematic region $0
\leq q^2 \leq (\mu_0 - \mu_m)^2$, where  $\omega\geq 1$.  It therefore
makes sense to redefine $\omega \equiv q_-/p_-$ rather than $p_-/q_-$,
so that $\omega \in [0,1]$ in this range.  One then finds
\begin{equation} \label{om2}
\omega (q^2) = \frac{1}{2} \left[ 1 + \left( \frac{q^2 -
\mu_m^2}{\mu_0^2} \right) -\sqrt{1- 2
\left( \frac{q^2 + \mu_m^2}{\mu_0^2} \right) + \left( \frac{q^2 
- \mu_m^2}{\mu_0^2} \right)^2} \, \right].
\end{equation}

	It is convenient to define the triple overlap integral
\begin{eqnarray} \label{f0nm}
\lefteqn{F_{0nm} (\omega) \equiv} & & \nonumber \\ & &
\left[ \frac{1}{1-\omega} \int_0^\omega dv \,
\phi_0^{1\bar 2} (v) \phi_n^{1\bar 3} \left(\frac v \omega \right)
\Phi_m^{3\bar 2} \left( \frac{v-\omega}{1-\omega} \right) - \frac 1
\omega \int^1_\omega dv \, \phi_0^{1\bar 2} (v) \Phi_n^{1 \bar 3}
\left( \frac v \omega \right) \phi_m^{3\bar 2} \left(
\frac{v-\omega}{1-\omega} \right) \right] ,
\end{eqnarray}
as well as the contact terms
\begin{eqnarray} \label{ct}
{\cal C}_1 & \equiv & -\frac{1}{\omega} \int_\omega^1 dv \,
\phi_0^{1\bar 2} (v) \phi_m^{3\bar 2} \left( \frac{v-\omega}{1-\omega}
\right) , \nonumber \\
{\cal C}_2 & \equiv & -\omega \int_\omega^1 dv \, \phi_0^{1\bar 2} (v)
\phi_m^{3\bar 2} \left( \frac{v-\omega}{1-\omega} \right)
\frac{1}{v (v- \omega)} , \nonumber \\
{\cal C}_3 & \equiv & -\frac{1}{1-\omega} \int_0^\omega dv \,
\phi_0^{1\bar 2} (v) \Phi_m^{3\bar 2} \left( \frac{v-\omega}{1-\omega}
\right) .
\end{eqnarray}
Note that the triple overlap is somewhat different from that in
Eq.~(\ref{fn0m}), both in the arguments of each wave function and the
definitions ((\ref{om1}) and (\ref{om2})) of $\omega$ for each case.
Furthermore, there is some flexibility in how one expresses results
containing these contact terms, since one can use the completeness of
the 't~Hooft model eigenfunctions on $x \in [0,1]$ to show $\sum_n c_n
\phi_n (x) = 1$, and from this prove identities such as
\begin{equation}
\sum_n c_n F_{0nm} (\omega) = {\cal C}_2 - {\cal C}_3 .
\end{equation}

	After a lengthy but straightforward calculation, one finds
\begin{eqnarray}
f_+ (q^2) & = & \sum_n \frac{{\cal A}_n (q^2)}{1-q^2/\mu_n^2} +
\frac{1}{(q^2/\omega - \mu_0^2 \omega)} \left\{ q^2 {\cal C}_1 -
\left[ 1 + (-1)^m m_1 m_3 \right] {\cal C}_2 + {\cal C}_3 \right\} ,
\end{eqnarray}
and
\begin{eqnarray}
f_- (q^2) & = & \frac{1}{q^2} \Biggl\{ \sum_n \frac{{\cal B}_n
(q^2)}{1-q^2/\mu_n^2} - \sum_n \frac{{\cal A}_n (q^2)}{1-q^2/\mu_n^2}
\left( \mu_0^2 - \mu_m^2 \right) \Biggr. \nonumber \\ & & 
+ \Biggl. (1-r) \left( q^2 {\cal C}_1 - {\cal C}_3 \right) - \left[
\left[ 1 + (-1)^{m+1} \right] - \left[ 1 + (-1)^m \right] m_1 m_3 r
\right] {\cal C}_2 \Biggr\} ,
\end{eqnarray}
where
\begin{equation}
r \equiv \frac{\mu_0^2 - \mu_m^2}{q^2/\omega - \mu_0^2 \omega} ,
\end{equation}
and the pole residue functions are given by (compare (\ref{fp1}) and
(\ref{fm1}))
\begin{equation}
{\cal A}_n (q^2) = \frac{c_n \left[ 1 + (-1)^{n+m} \right]}{\left(
q^2/\omega - \mu_0^2 \omega \right)} F_{0nm} (\omega) ,
\end{equation}
and
\begin{equation}
{\cal B}_n (q^2) = c_n \left[ 1 + (-1)^{n+m+1} \right] F_{0nm}
(\omega) .
\end{equation}
Finally, the matrix element for the decay {\bf 0} $\to$ {\bf mk},
which unlike Eq.~(\ref{mt1}) holds for all such decays allowed by
kinematics, is given by
\begin{eqnarray} \label{mt2}
{\cal M}_T & = & c_{k} \sqrt{\frac{2}{\pi}} \frac{G_F M_W^2}{(M_W^2 -
q^2)} V_{31} V^*_{45} \nonumber \\ & &
\cdot \left[ 2 (c_V^2 - c_A^2) \left\{ \sum_n \frac{\left[
(-1)^{k} q^2 + (-1)^n \mu_n^2 \right] c_n}{q^2 - \mu_n^2} F_{0nm}
(\omega) \nonumber + (-1)^{k+1} q^2 {\cal C}_1 + m_1 m_3 {\cal C}_2
\right\} \right. \nonumber \\ & &
\hspace{1em} +\frac{q^2}{M_W^2} \left[ (c_V + c_A) (-1)^{k} - (c_V -
c_A) \right] \nonumber \\ & & 
\hspace{2em} \cdot \Biggl\{ \sum_n \frac{c_n}{q^2 - \mu_n^2} \left[
(c_V + c_A) (-1)^n \mu_n^2 - (c_V - c_A) q^2 \right] F_{0nm} (\omega)
\Biggr. \nonumber \\ & &
\hspace{3em} \left. \Biggl. + (c_V - c_A) q^2 {\cal C}_1 + (c_V + c_A)
m_1 m_3 {\cal C}_2 \Biggr\} \right] ,
\end{eqnarray}
where as before, $q^2 = \mu_{k}^2$.

\section{Results and Discussion} \label{res}

	We computed the weak decay width of a free heavy quark with
masses $M = 2.28 \to 15.00$ in units of $g \sqrt{N_c /2\pi}$ using
Eq.~(\ref{partw1}), the $M_W \to \infty$ case.  Likewise, we computed
the hadronic width using the same range of heavy quark mass and a
fixed light quark mass $m = 0.56$.  The expressions used were
Eqs.~(\ref{mt2}) and (\ref{two}), with definitions (\ref{cn}),
(\ref{om2}), (\ref{f0nm}), (\ref{ct}), and with sums over all channels
${\bf 0} \to {\bf mk}$ satisfying the on-shell condition $\mu_m +
\mu_{k} \leq \mu_0$.  Both widths are taken to have the same overall
multiplicative factor $2 G_F^2 |V_{31} V_{45}^*|^2 (c_V^2 - c_A^2)^2
/\pi$.

	It is equally possible, in principle, to use (\ref{mt1})
instead of (\ref{mt2}) and backsolve for pole residues ${\cal A}_n$
and ${\cal B}_n$ defined in (\ref{an1})--(\ref{bn1}), or equivalently
the overlap integrals $F_{n0m}$, using the expressions in Appendix B
whenever $\mu_n < \mu_0 + \mu_m$, and obtain the hadronic width in
this way.  However, as discussed in Sec.~\ref{exc}, this approach
rapidly leads to uncontrollably large numerical uncertainties.
Nevertheless, we were able to show in some simple cases with only a
few backsolved residues that both methods produce the same numerical
result within a few percent.

	It is no more difficult to consider cases other than $M_W \to
\infty$.  For example, if one imposes the condition $V-A$ condition
$c_V = c_A = 1/2$, then Eq.~(\ref{mt2}) is just as valid, but now one
uses the partonic width (\ref{partw2}).

	Of course, the partonic width is just a single easily
evaluated function of the quark masses.  The hadronic width, on the
other hand, requires first the solution of the 't~Hooft equation,
which is accomplished by means of the Multhopp technique described in
Appendix A, repeated for as many resonances as desired.  Next, the
matrix elements are obtained by taking sums of overlap integrals over
these wave functions, as in Eqs.~(\ref{f0nm}) and (\ref{ct}).  We
compute the first 500 eigenvectors but include only the first 50 in
our sums over resonances. The results change very little when more
resonances are included.  Finally, the amplitude for a given exclusive
process is squared and multiplied by phase space to give the hadronic
width.

	Clearly, such a procedure uses a significant amount of
computing time, and therefore it is not practical to compute the
hadronic width at points exceptionally finely spaced in $M$.  In
practice, we computed each above-threshold amplitude at values of
$M=2.28$ and each integer mass from $M=3.00$ to 15.00.  The
significance of the lower bound is that, with the given light quark
mass $m = 0.56$, this ``$b$ quark'' mass gives a ground-state ``$\bar
B$'' meson just above the threshold for producing two ground-state
``$\pi$'' mesons, {\it i.e.}, the smallest value of heavy quark mass
unstable under hadronic weak decay.

	We then make the empirical observation that the amplitudes for
exclusive processes ${\cal M}_T ({\bf 0} \to {\bf mk})$ are smooth
functions of $M$.  We thus obtain the value of the amplitude at all
intermediate points $M$ by fitting to a fixed power law behavior over
each interval, either by interpolating for values between adjacent
pairs of points where the amplitudes were computed directly, or by
extrapolating from the nearest two points if we are probing values of
$M$ above the process threshold but below the first explicitly
computed point.  In fact, we find that the exclusive amplitudes do not
vanish at threshold and are usually\footnote{In the few exceptions to
this rule, the amplitude dips slightly for values of $M$ just above
threshold, but thenceforth assumes monotonically increasing behavior.}
monotonically increasing functions of $M$ (for example, see Fig.~3),
although the rate of this increase is dependent upon the particular
exclusive mode under consideration.  The phase space is a known
function of the computed meson masses, and thus the width can be
reliably computed at any value of $M$ in the desired range.

\epsfxsize 6.0 truein
\INSERTFIG{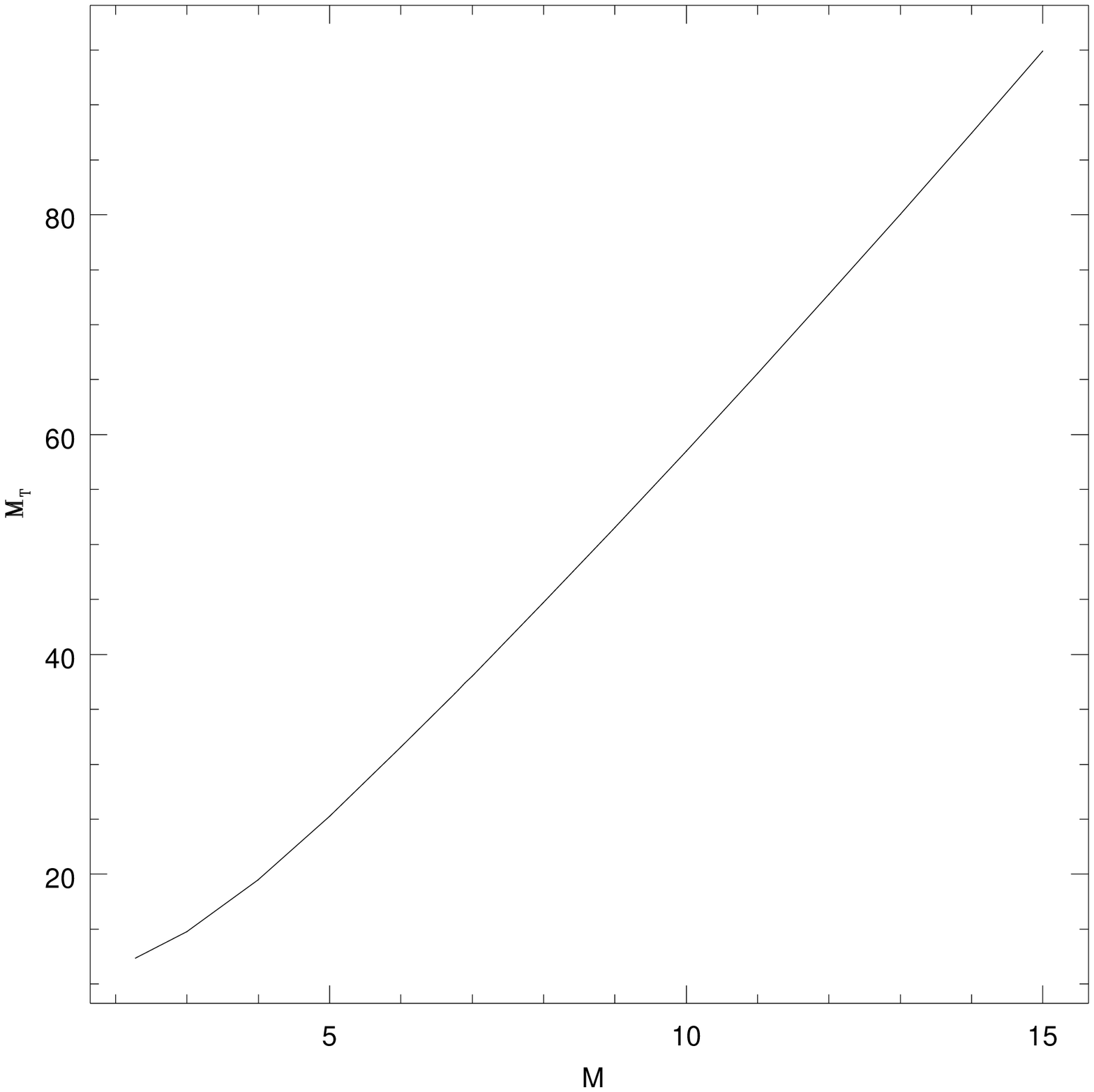}{3}{Weak decay amplitude ${\cal M}_T$ for the
exclusive decay to the lowest mode ${\bf 0} \to ({\bf m} = 0), ({\bf
k} = 0)$, as a function of heavy quark mass $M$, with light quark mass
$m = 0.56$.  The overall factor $2\sqrt{2/\pi} \, G_F V_{31} V_{45}^*
(c_V^2 - c_A^2)$ in the amplitude is suppressed for
convenience.\hfill}

	Since the phase space in 1+1 is singular at threshold
(Eq.~(\ref{two})), one would expect a plot of width $\Gamma$ {\it vs.\
}$M$ to be very ill-behaved, with dramatic singularities increasing in
density as $M$ increases.  One would expect it to be essential to use
some sort of smearing in $M$ to properly test duality between this
hadronic description of the $\Gamma$ and the smooth partonic result.
In fact, this does not appear to be the case.  We refer to Fig.~4,
which is our central result.  It is obtained by interpolating each
exclusive decay amplitude, as described above, at intervals of $\Delta
M = 0.01$.  The remarkable result is that, after passing the first
couple of thresholds, $\Gamma$ appears to be a nearly smooth function
in $M$, barely sensitive to the phase space singularities as each new
threshold is passed.  This result suggests that the effect of
individual higher resonances is quite minimal, as one might expect in
3+1 dimensions.  In the 1+1 case, however, the result is all the more
surprising, since now phase space near threshold provides a large
enhancement rather than a suppression.

\epsfxsize 6.0 truein
\INSERTFIG{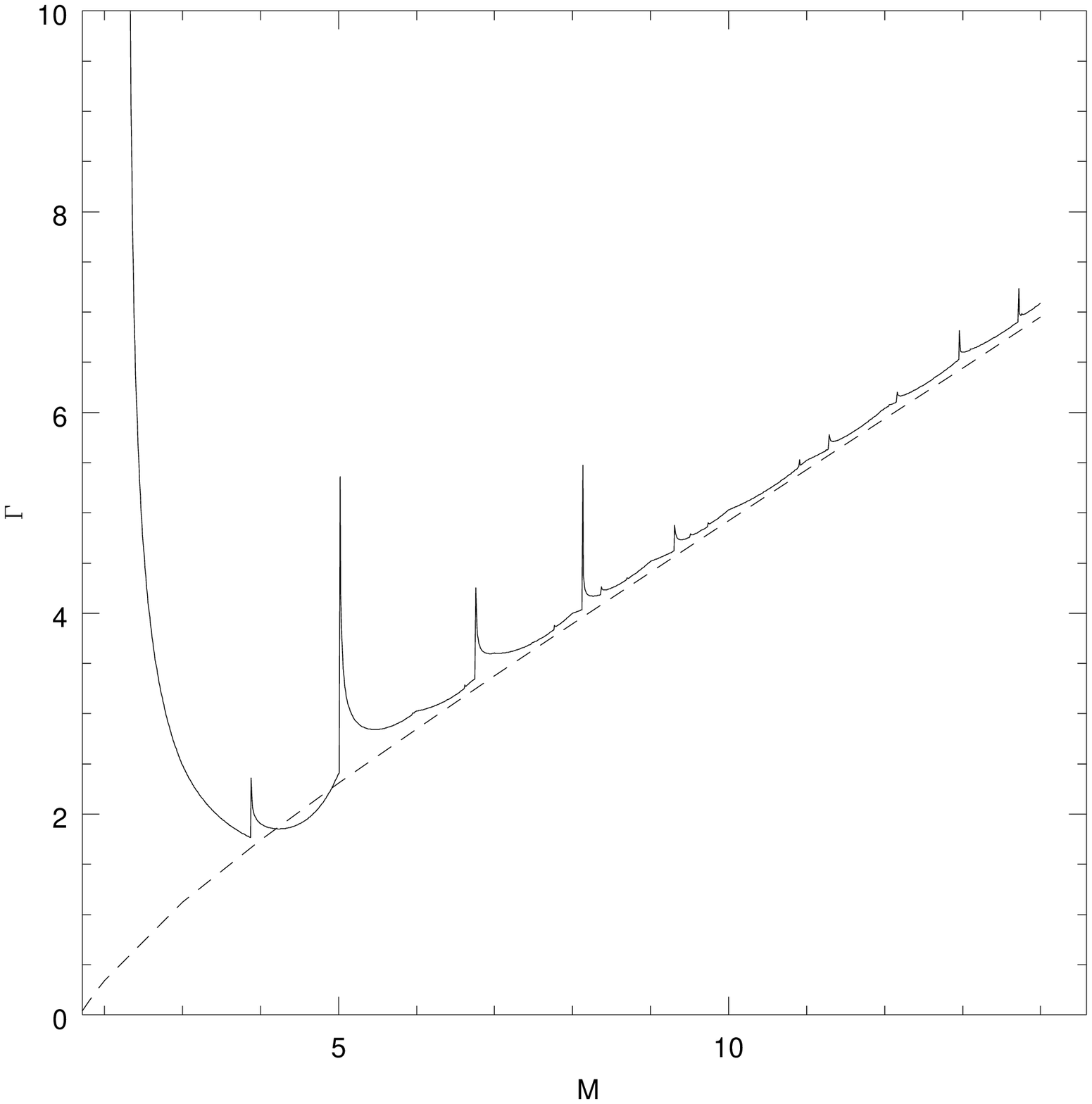}{4}{The full decay width for the sum of
exclusive modes in the decay ${\bf 0} \to {\bf mk}$ as a function of
heavy quark mass $M$, with light quark mass $m = 0.56$.  The overall
factor $8 G_F^2 |V_{31} V_{45}^*|^2 (c_V^2 - c_A^2)^2 /\pi$ in the
width is suppressed for convenience.  The dashed line is the
tree-level parton result, Eq.~(\ref{partw1}).}

	It is interesting to watch the width develop as more and more
resonances are added.  In Figs.~5$a$--5$d$ we include exclusive
channels with the lowest 1, 3, 5, and 11 thresholds, respectively.  We
now see explicitly that the full width over the range in $M$ we
consider is essentially produced by the first 11 channels, indicating
the decreasing influence of individual higher resonances.  The small
wave in $\Gamma$ above all the included thresholds is an artifact due
to the interpolation routine between values of $M$ at which the
amplitudes are explicitly computed; its small size indicates the
smoothness of the amplitudes in $M$ and the reliability of the
interpolation.\\
\epsfxsize 6.0 truein
\INSERTFIG{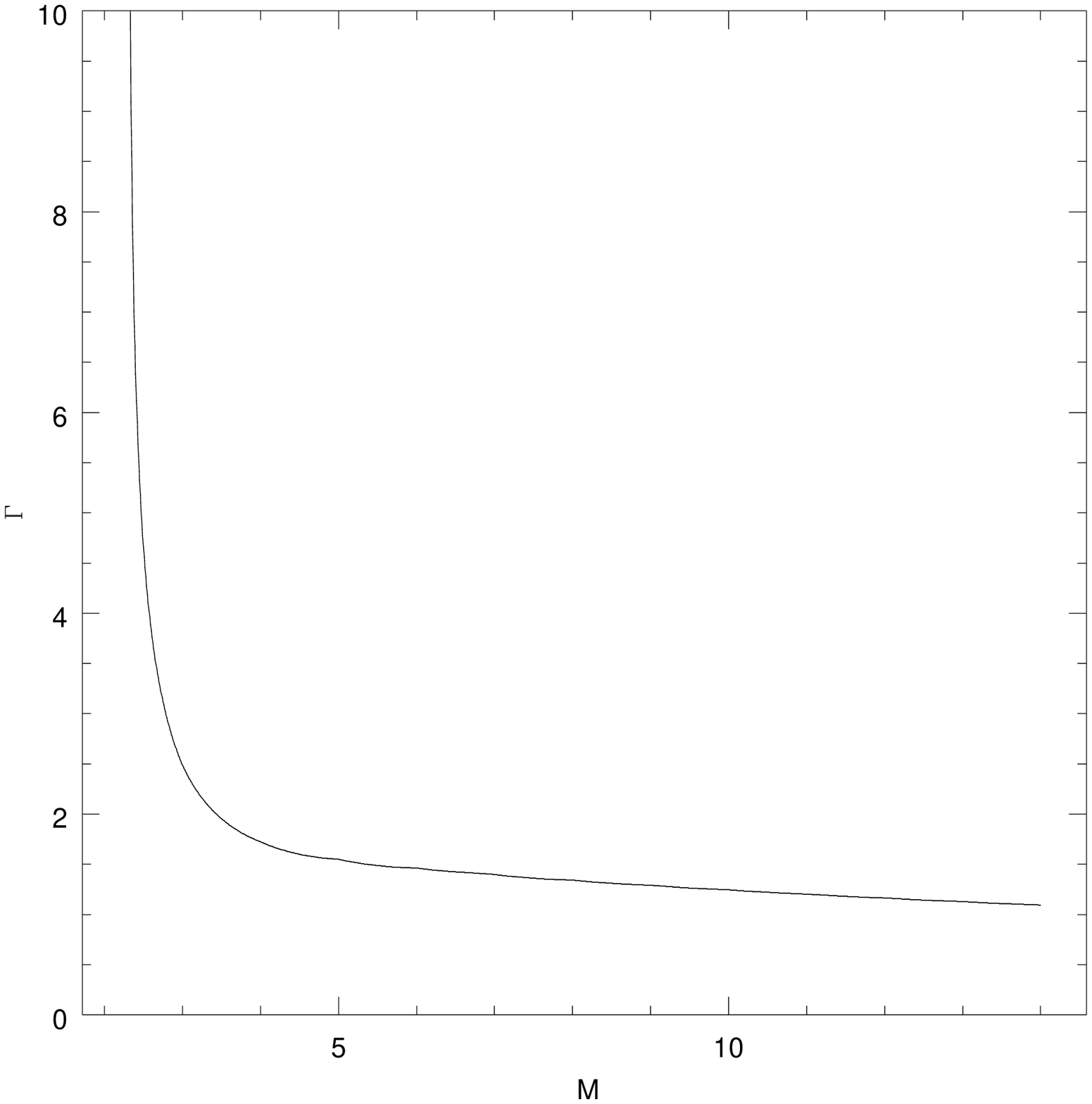}{5$a$}{The full decay width as a function of
heavy quark mass $M$, with light quark mass $m = 0.56$, including only
the exclusive mode with the lowest threshold value (corresponding to
${\bf m} = {\bf k} = 0$).  The scale is the same as in Fig.~4.}
\epsfxsize 6.0 truein
\INSERTFIG{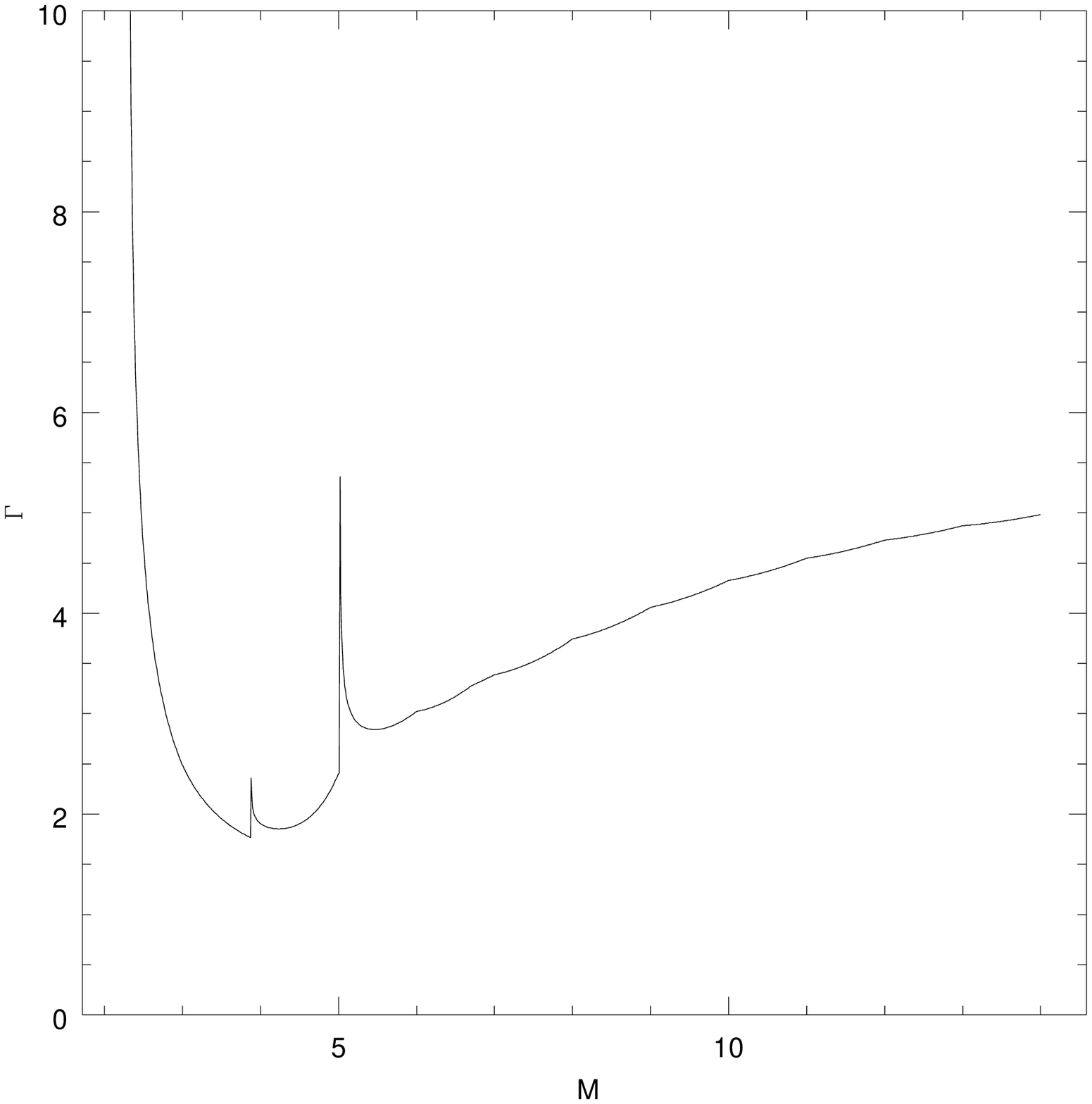}{5$b$}{Same as Fig.~5$a$, except now including
the exclusive modes corresponding to the {\rm three} lowest threshold
values.}
\epsfxsize 6.0 truein
\INSERTFIG{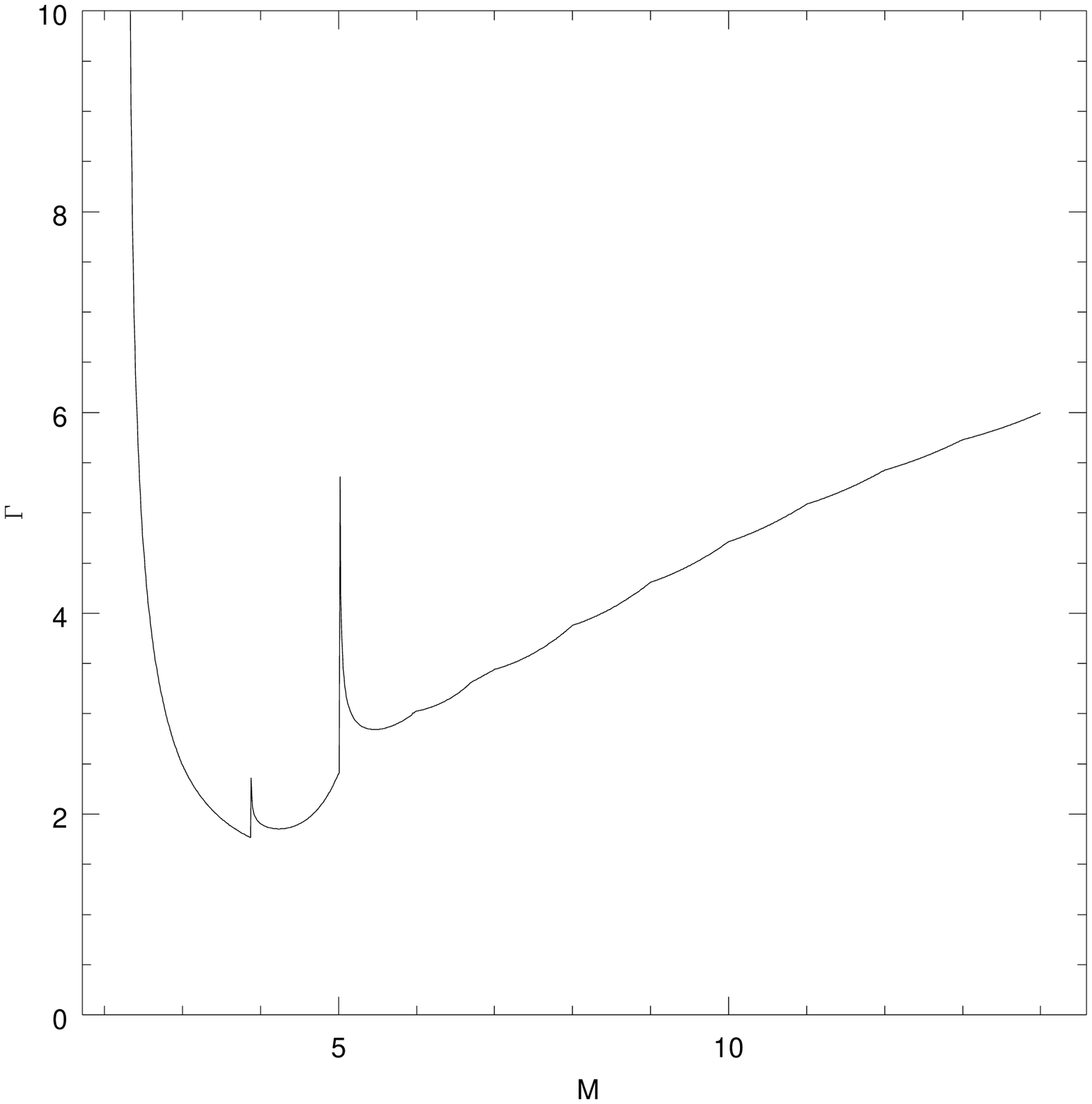}{5$c$}{Same as Fig.~5$a$, except now including
the exclusive modes corresponding to the {\rm five} lowest threshold
values.}
\epsfxsize 6.0 truein
\INSERTFIG{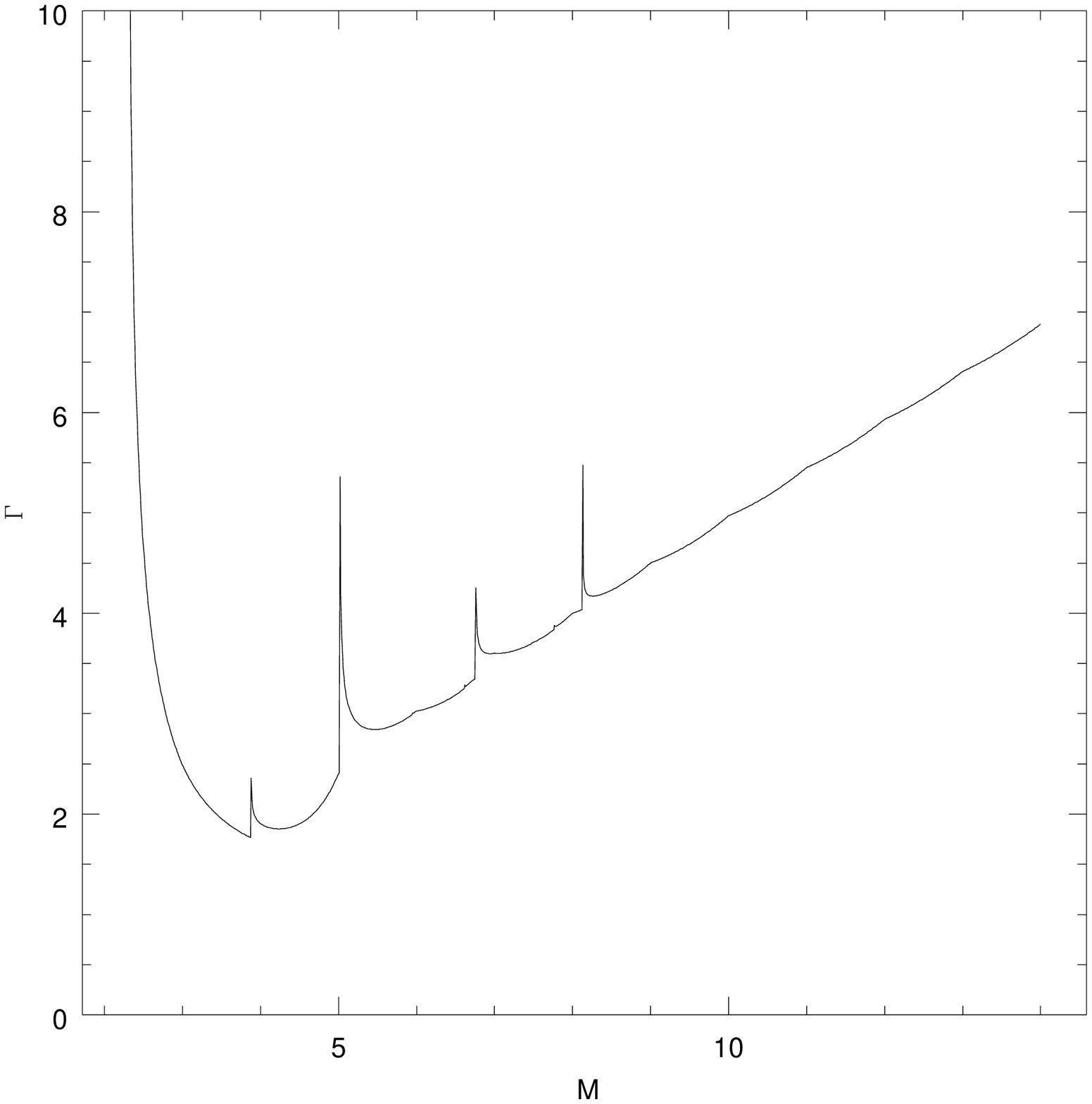}{5$d$}{Same as Fig.~5$a$, except now including
the exclusive modes corresponding to the {\rm eleven} lowest threshold
values.  Observe that this figure is almost indistinguishable from the
full result, Fig.~4.}

	Another remarkable feature of Fig.~4 is the near-perfect
linearity of $\Gamma$ for values $M > 7.0$.  Suppressing the
proportionality constant between $\Gamma$ and $M$, Fig.~4 appears to
obey the asymptotic form $\Gamma \approx 0.514 M -0.141$.  This is
surprisingly close to what is predicted asymptotically for the partonic
rate: Suppressing the same proportionality constant in
Eq.~(\ref{lim1}), one predicts $\Gamma_{\rm part} =\frac12 M(1+O(1/M^2))$.

	One may ask whether the strength of the peaks in Fig.~4 is
large enough that the mass-smeared partonic and hadronic widths
nevertheless disagree.  That is, local duality appears remarkably well
satisfied, but perhaps global duality actually fails by concealing a
large portion of $\Gamma(M)$ in the very narrow threshold peaks.  In
this scenario, the apparent asymptotic smoothness of Fig.~4 fools us,
for the density of threshold singularities increases with $M$ so
rapidly as to push the curve of hadronic $\Gamma (M)$ out of agreement
with $\Gamma_{\rm part} (M)$ for sufficiently large $M$.  We
now argue that this possibility does not appear to be realized, at
least numerically.  Let us smear in $M$ over a region of
size~$\Delta$, $1\ll\Delta\ll M$.\footnote{In practice, we use a
normalized Gaussian window function with a mean of $M$ and a variation
of $\Delta$, although the result should be independent of the
particular form used.}  Owing to the approximate linearity of squared
meson masses in the excitation number, there are $\sim M^3\Delta$
thresholds in this region, and the contribution to the smeared rate
from their phase space near the threshold scales as $M^{-5/2}$ for
each.\footnote{This is verified from (\ref{two}) and the observation
that, for $M \gg 1$, $\mu_0 \propto M$.}  We observe empirically that
the magnitudes of amplitudes first appearing at a given threshold mass
$M_{\rm thr}$ tend to evolve approximately no faster than as $M_{\rm
thr}^{-0.6}$.  It follows that the contribution to the smeared $\Gamma
(M)$ from the region of width $\Delta$ scales approximately as
$M^{-0.7}$.  For the border region, where one of the mesons $m,k$ is
highly excited and the other is near the ground state, the phase space
is seen to scale as $M^{-2}$, but the number of such states is only
$\sim M\Delta$, so again the area under these peaks contributes little
to $\Gamma(M)$.  Finally, the phase space far above a given threshold
scales as $M^{-3}$, which means that, given the density of states for
various eigenvalue indices, amplitudes cannot on the average grow with
$M$ above their respective thresholds faster than $M^{1/2}$ for $k \gg
1$ and $m \gg 1$, $M^{3/2}$ for one of $m,k = O(1)$ and the other $\gg
1$, or $M^2$ for $m$ and $k = O(1)$, or else the linear behavior of
$\Gamma(M)$ will be violated.  In fact, the amplitudes we have
computed all obey these constraints.  We see that the linear behavior
observed requires a delicate balance of numbers and mass dependences
of amplitudes versus excitation numbers, and we hope to obtain
analytic arguments for this remarkable behavior in the future.

	What are we to conclude from this result?  The 't Hooft model
is exactly soluble, so it must be the case that the fully-dressed
parton diagrams give results agreeing with the hadronic calculation;
indeed, this is how the hadronic problem was solved in the first
place.  The partonic width computed in (\ref{partw1}) represents only
the Born term in an expansion in strong coupling $g$, so the addition
of gluon loops is apparently necessary to bring the two results into
agreement.  The small discrepancy between the curves may have this
origin, or it may simply be a limitation of the numerical accuracy of
the calculation.  However, it is interesting to note that the two
curves appear to differ asymptotically by a constant, which for plots
linear in $M$ is a $1/M$ correction.  Therefore, we suggest that this
effect is genuine and not a numerical artifact.  In Fig.~6 we
superimpose on the hadronic width of Fig.~4 the curve $\Gamma_{\rm
part}(M)\cdot (1+0.15/M)$, and see that the fit is outstanding.  From this
result, we learn that local duality for this system is violated badly
only for the first few resonances and very close to thresholds of
higher resonances, and that $1/M$ effects appear to be only at the few
percent level for $M>7$.  It would be very interesting to see
explicitly what happens to the partonic inclusive width at the one- or
two-loop level.
\epsfxsize 6.0 truein
\INSERTFIG{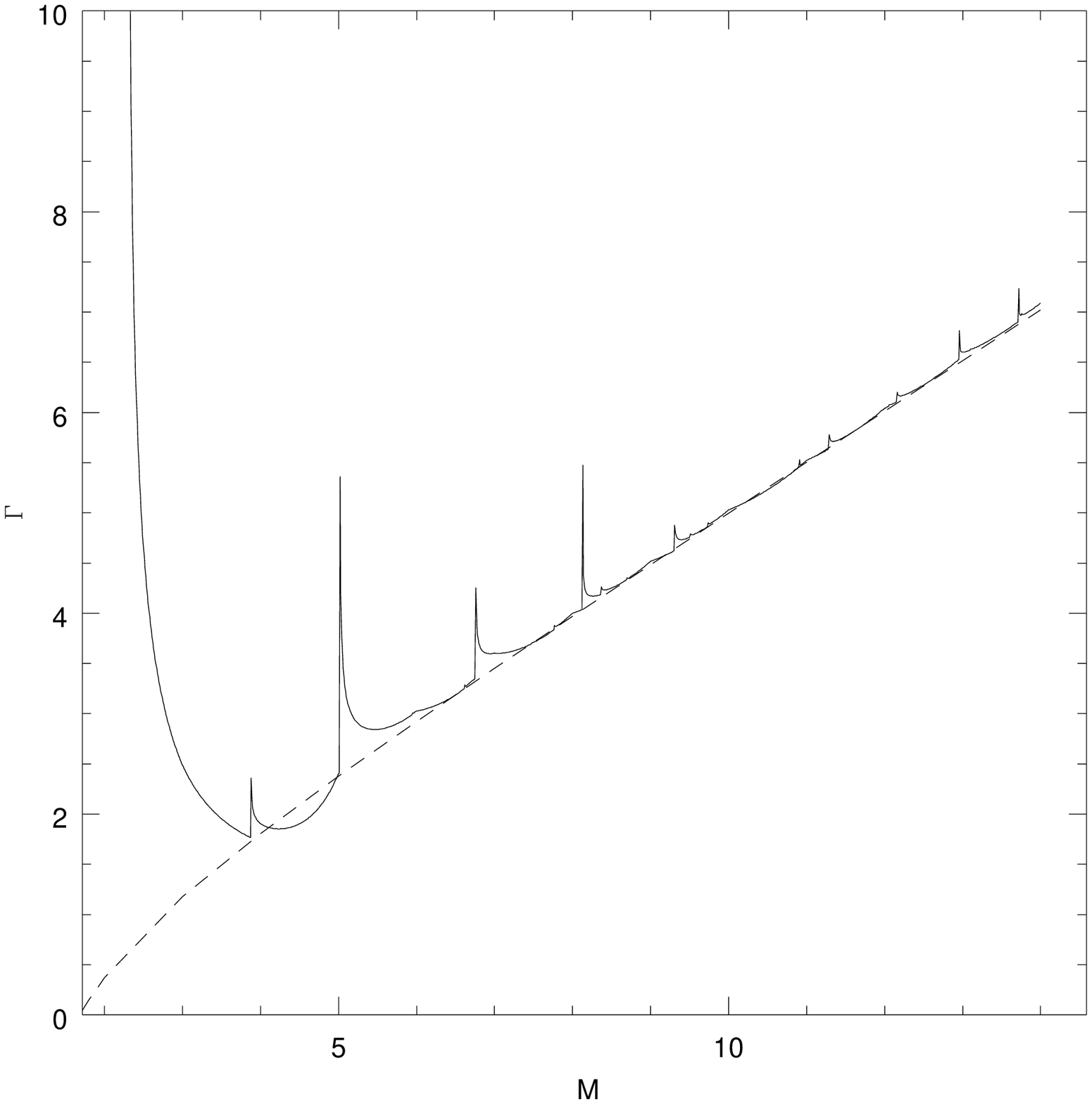}{6}{The full decay width of Fig.~4 compared to
the tree-level parton result of Eq.~(\ref{partw1}) corrected by a
$1/M$ effect: $\Gamma_{\rm part}(M)\cdot(1+0.15/M)$.}

	One natural idea of how to improve the Born result is to
replace the bare quark masses with the renormalized values.  This
would not sum all gluon corrections, but it would include an important
subclass of them.  Unfortunately, due to the result (\ref{mren}),
masses below 1.0 (such as that of our light antiquark) have {\em
imaginary\/} renormalized values, and then our whole interpretation of
phase space, essential for the calculation of the width, becomes
ambiguous.

\section{Conclusions} \label{conc}

	We have calculated the nonleptonic decay width of a
heavy-light meson in the context of the 't~Hooft model as a function
of the bare heavy quark mass, both for the Born term of the free
partonic decay (which we called the ``partonic width'') and the full
sum of allowed hadronic decays (the ``hadronic width'').  We found
that these two quantities approximately agree at leading order in $M$,
with the hadronic width being slightly larger.  Both quantities are
observed to grow linearly and smoothly for large $M$, despite the
effects of numerous phase space threshold singularities in the
hadronic case. The slight discrepancy between hadronic and partonic
widths is well-fit by a $1/M$ correction, $\Gamma_{\rm
hadr}(M)\approx\Gamma_{\rm part}(M)\cdot(1+0.15/M)$.

	Assuming that the small discrepancy between the partonic and
hadronic results is genuine (rather than a numerical artifact) leads
one to conclude that nonleptonic heavy-light meson decays in 1+1
dimensions cannot be described in terms of an OPE that lacks $1/M$
corrections, and it naturally leads one to believe that the same
conclusion is true in 3+1.  Since the lowest order of the OPE is
simply the naive free quark picture, this result also has obvious
implications for the application of quark models in such decays.
Another incisive test of quark-hadron duality in 1+1 is whether
annihilation diagrams, in which the valence quarks in the decaying
meson annihilate through a weak current, are suppressed compared to
spectator tree diagrams (Fig.~2); these studies are well
underway\cite{GLII}, and results will be forthcoming shortly.

	A number of unanswered questions not addressed by this work
include the effects of loop corrections to the Born amplitude free
quark decay, the dependence of decay widths on the light quark mass,
the effects of including finite meson strong decay widths (which are
$O(1/N_c)$), the effects of identical final state quarks or mesons,
multiparticle final states (also suppressed by powers of $N_c$), and
so on.  While ``two-dimensional phenomenology'' cannot be used as a
quantitative substitute for the standard four-dimensional variety, it
clearly indicates the limitations of the standard lore.

	{\it Note Added}.  An interesting recent work by
Blok\cite{blok} suggests that global quark-hadron duality at high
energies in the 't~Hooft model with massless quarks may be achieved by
including smearing through the $1/N_c$-suppressed widths of
resonances.  Our calculation, on the other hand, does not include
finite-width effects but nevertheless achieves an effectively smeared
result, even at relatively low $M$, which supports the claim of
duality at leading order.

\vskip1.2cm
{\it Acknowledgments}
\hfil\break
This work is supported by the Department of Energy under contract
DOE-FG03-97ER40506.

\appendix
\section{The Multhopp Technique}

	This technique\cite{JM,Mul} is used to solve numerically
certain singular integral equations in a systematic expansion of basis
functions.  Specifically, it is used to solve equations of the form
\begin{equation}
\psi(x) = \int_a^b dy \, K(x,y) \, \Pr \frac{1}{(x-y)} \cdot
\frac{d \psi(y)}{d y} ,
\end{equation}
where $\psi(a) = \psi(b) = 0$.  The 't~Hooft equation is seen to be of
this form after an integration by parts.  Generally speaking, one maps
the interval $y \in [a,b]$ to $\theta \in [0,\pi]$ by a function
linear in $\cos \theta$, and then expands in a Fourier series in
$\theta$.  Equivalently, such functions written directly in terms of
$y$ for each mode turn out to be the product of a common factor
$\left[\sqrt{y(1-y)} \, \right]$ times Chebyshev polynomials of the
second kind, $U_n(y)$.

	In particular, we use a slight variant of the Multhopp
technique for the special case that $K(x,y)$ is independent of $y$.
The variable transformation is
\begin{equation}
x = \frac{1+\cos \theta}{2}, \, \, y = \frac{1+\cos \theta'}{2} ,
\end{equation}
in terms of which the 't~Hooft equation may be written
\begin{equation}
\frac{\mu^2}{2} \phi^{M\overline{m}} (\theta) = \frac{(M^2+m^2) -
(M^2-m^2)\cos \theta}{\sin^2 \theta} \phi^{M \overline{m}} (\theta) +
\int_0^\pi d\theta' \, \frac{d \phi^{M\overline{m}} (\theta')}{d
\theta'} \, \Pr \frac{1}{(\cos \theta - \cos \theta')} .
\end{equation}
Expanding
\begin{equation} \label{exp}
\phi^{M\overline{m}} (\theta) = \sum_{m=1}^{\infty} a_m \sin m\theta ,
\end{equation}
and using the integral ($m = 0,1,\ldots$)
\begin{equation}
\int_0^\pi d\theta' \, \Pr \frac{1}{(\cos \theta' - \cos \theta)}
\cos m\theta' = \pi \frac{\sin m\theta}{\sin \theta} ,
\end{equation}
we are led to a series equation for the
eigenvector coefficients $a_m$.

	We truncate the series after mode $m=N$, and evaluate both
sides at the equally-spaced values of $\theta$ ({\em Multhopp
angles}),
\begin{equation}
\theta_k \equiv \frac{k \pi}{N+1}, \hspace{1em} \, m =
1\ldots N ,
\end{equation}
which are a convenient choice because of the inversion identity
\begin{equation}
\sum_{k=1}^N \sin k\theta_m \sin k \theta_n = \frac{1}{2} (N+1)
\delta_{mn} ,
\end{equation}
to obtain at last the finite eigenvector system (compare
Ref.~\cite{JM}, Eqs.~(A.6)--(A.7))
\begin{equation}
\mu^2 a_n = \frac{4}{(N+1)} \sum_{m=1}^N \sum_{k=1}^N \frac{\sin
\theta_{kn} \sin \theta_{km}}{\sin \theta_k} \left\{ \frac{(M^2+m^2) -
(M^2-m^2) \cos \theta_k}{\sin \theta_k} + m \pi \right\} a_m .
\end{equation}

	The series expansion (\ref{exp}) transforms the normalization
condition $\int_0^1 dx \, \phi(x)^2 = 1$ into
\begin{equation}
1 = - \sum_{m=1}^N m a_m \sum_{n=1}^N n a_n \left[ 1 +
(-1)^{m+n} \right] \left[ \left( 1 - (m-n)^2 \right) \left( 1 -
(m+n)^2 \right) \right]^{-1} ,
\end{equation}
while the phase of the eigenvectors may be chosen by noting the
asymptotic forms
\begin{equation}
\phi(x) \to -2 \sqrt{x} \sum_{m=1}^N (-1)^m m a_m + O(x),
\end{equation}
as $x \to 0$, and
\begin{equation}
\phi(x) \to +2 \sqrt{1-x} \sum_{m=1}^N m a_m + O(1-x), 
\end{equation}
as $x \to 1$.  Note that the Multhopp solution requires the
eigenfunctions to vanish as square roots at the endpoints, in contrast
to the dynamically-generated exponents of the exact solution given by
Eq.~(\ref{asymp}).  Great care must be exercised when extracting
information near the endpoints of the numerically-calculated wave
functions.

	In terms of the eigenvector components $a_m$, the vertex
function $\Phi$ may be written for real $x$,
\begin{equation}
\Phi (x<0) = \frac{-\pi}{\sqrt{x(x-1)}} \sum_{m=1}^N (-1)^m m a_m
\left( \sqrt{1-x} - \sqrt{-x} \right)^{2m} ,
\end{equation}
and
\begin{equation}
\Phi (x>1) = \frac{+\pi}{\sqrt{x(x-1)}} \sum_{m=1}^N m a_m
\left( \sqrt{+x} - \sqrt{x-1} \right)^{2m} .
\end{equation}
These regions are the only ones for which we need explicit
expressions; whenever $x \in (0,1)$, the 't~Hooft equation (\ref{tHe})
and the definition (\ref{bphi}) may be employed to rewrite $\Phi(x)$
in terms of $\phi(x)$.

	Because of the singularities in $\Phi$ at $x=0$ and 1, when
numerically computing overlap integrals such as in Eqs.~(\ref{fn0m}),
(\ref{f0nm}), or (\ref{ct}), one must sample the region near the
singularities more heavily.  Moreover, in contrast to the numerical
Multhopp solution for $\Phi(x)$ given above, the singularities in the
exact solution behave as $\Phi^{M \overline{m}} (x \to 0^-) \sim
(-x)^{\beta_M-1}$ and $\Phi^{M
\overline{m}} (x \to 1^+) \sim (1-x)^{\beta_m-1}$.  Such a
difference can also lead to substantial errors when computing overlap
integrals; the answer is to increase the number of eigenvector modes
in the wave function solution, so as to decrease the size of the
region where the asymptotic behaviors differ.

	Finally, the quantity $c_n$, which is the meson decay constant
up to a normalization factor (Eq.~(\ref{dec})), is simply given by
\begin{equation}
c_n \equiv \int_0^1 dx \, \phi_n(x) = \frac{\pi}{4} a_1.
\end{equation}

\section{Amplitude Sum Rules}

	The asymptotic forms of 't~Hooft model solutions may be used
to obtain constraints on the form factor expressions, such as
Eqs.~(\ref{fp1})--(\ref{fm1}), that are derived from them, as is shown
in \cite{Ein}.  There it is seen that, as $|q^2| \to \infty$, the form
factors vanish at least as fast as $|q^2|^{1+\beta_m}$, where
$\beta_m$ is defined in Eq.~(\ref{asymp}).  By symmetry there is also
a term that falls off as $|q^2|^{1+\beta_M}$, but since $M > m$, we
have $\beta_M > \beta_m$, and the slower fall-off dominates.

	It is then apparent from Eqs.~(\ref{fp1})--(\ref{fm1}) that
the pole residue functions ${\cal A}_n (q^2)$ and ${\cal B}_n
(q^2)/q^2$ must also vanish at least as fast as $|q^2|^{\beta_m}$.
But (\ref{fp1}) and (\ref{fm1}) have a very suggestive form: They
explicitly display the pole-dominance nature of the large-$N_c$ limit,
since ${\cal A}_n (q^2)$ and ${\cal B}_n (q^2)/q^2$ are expected to
have no non-analytic behavior for finite $q^2$, and moreover, these
expressions are written very conveniently for an application of
Cauchy's theorem.  If we choose our contour such that it encloses all
of the complex $q^2$ plane except for the part of the real axis where
the resonance poles lie, then using Cauchy's theorem with the
vanishing of the residue functions for $|q^2| \to
\infty$ gives\cite{Ein}
\begin{equation} \label{srule}
\sum_n \frac{\Lambda_n (q^2)}{1- q^2/\mu_n^2} = 
\sum_n \frac{\Lambda_n (\mu_n^2)}{1- q^2/\mu_n^2} ,
\end{equation}
where $\Lambda_n (q^2) = {\cal A}_n (q^2)$ or ${\cal B}_n (q^2)/q^2$,
and the sums over $n$ are restricted to the appropriate parities.
Each residue function thus needs only to be computed at the value of
its corresponding mass eigenvalue.

	This sum rule also gives us enough information, in principle,
to obtain residue functions that are not conveniently computable in a
direct fashion\cite{JM}.  Let $t$ denote some threshold in $n$, below
which it is unwieldy to compute the residue functions directly.  Then,
if we choose at least $t$ values of $q_i^2$, $i = 1\ldots t\/$ above
threshold, we obtain a solvable $t \times t$ linear system of
equations:
\begin{equation}
\sum_{n \leq t} \frac{\Lambda_n (\mu_n^2)}{1- q_i^2/\mu_n^2} \, = \,
-\sum_{n > t} \frac{\Lambda_n (\mu_n^2)}{1- q_i^2/\mu_n^2} \,
+ \, \sum_{n} \frac{\Lambda_n (q_i^2)}{1- q_i^2/\mu_n^2} .
\end{equation}
Adding additional $q_i^2$ simply overconstrains the system and
provides consistency checks.  We refer to this approach in the text as
``backsolving'', to indicate that it is an indirect method of
solution.

	We also note one other sum rule\cite{Ein,JM}, which is simply
derived by taking the limit $q^2 \to \infty$ in (\ref{srule}) and
recalling that the form factor expressions vanish in this limit:
\begin{equation}
\sum_n \mu_n^2 \Lambda_n (\mu_n^2) = 0.
\end{equation}

\newpage
\begin{center}
{\bf Figure Captions}
\end{center}

FIG.\ 1. Parton decay diagram for the inclusive decay.  Of interest
are the parton labels, as used in the text.

FIG.\ 2. Diagram for ``tree'' (T) meson exclusive decay.  Numbers
indicate quark labels used in the text (except {\bf 0}, which refers
to the ground-state ``$\bar B$'' meson), while letters indicate the
eigenvalue index of meson resonances.  One can also consider
contact-type diagrams, in which the point labeled by\/ {\bf n} is not
coupled to a resonance.

FIG.\ 3. Weak decay amplitude ${\cal M}_T$ for the exclusive decay to
the lowest mode ${\bf 0} \to ({\bf m} = 0), ({\bf k} = 0)$, as a
function of heavy quark mass $M$, with light quark mass $m = 0.56$.
The overall factor $2\sqrt{2/\pi} \, G_F V_{31} V_{45}^* (c_V^2 -
c_A^2)$ in the amplitude is suppressed for convenience.

FIG.\ 4. The full decay width for the sum of exclusive modes in the
decay ${\bf 0} \to {\bf mk}$ as a function of heavy quark mass $M$,
with light quark mass $m = 0.56$.  The overall factor $8 G_F^2 |V_{31}
V_{45}^*|^2 (c_V^2 - c_A^2)^2 /\pi$ in the width is suppressed for
convenience.  The dashed line is the tree-level parton result,
Eq.~(\ref{partw1}).

FIG.\ 5$a$. The full decay width as a function of heavy quark mass
$M$, with light quark mass $m = 0.56$, including only the exclusive
mode with the lowest threshold value (corresponding to {\bf m} = {\bf
k} = 0).  The scale is the same as in Fig.~4.

FIG.\ 5$b$. Same as Fig.~5$a$, except now including the exclusive
modes corresponding to the {\em three} lowest threshold values.

FIG.\ 5$c$. Same as Fig.~5$a$, except now including the exclusive
modes corresponding to the {\em five} lowest threshold values.

FIG.\ 5$d$. Same as Fig.~5$a$, except now including the exclusive
modes corresponding to the {\em eleven} lowest threshold values.
Observe that this figure is almost indistinguishable from the full
result, Fig.~4.

FIG.\ 6. The full decay width of Fig.~4 compared to
the tree-level parton result of Eq.~(\ref{partw1}) corrected by a
$1/M$ effect: $\Gamma_{\rm part}(M)\cdot(1+0.15/M)$.

\end{document}